\documentclass[%
 article,
 jmp,%
 amsmath,amssymb,
preprint,50pt%
]{revtex4-1}

\usepackage{amsfonts}
\usepackage{bm}%
\usepackage[colorlinks=true,linkcolor=blue,filecolor=blue,menucolor=yellow,urlcolor=blue,
citecolor=blue,anchorcolor=blue]{hyperref}
\usepackage{graphicx}% Include figure files
\usepackage{amsmath}
\usepackage{siunitx}
\makeatletter
\newcommand*{\rom}[1]{\expandafter\@slowromancap\romannumeral #1@}
\makeatother
\usepackage{rotating}
\usepackage{url}
\usepackage[utf8]{inputenc}
\usepackage{xr}
\usepackage{cleveref}
\usepackage{ccaption}
\usepackage{floatrow}
\usepackage[normalem]{ulem}
\usepackage[font= large,label font=bf,labelformat=simple]{subfig}

\usepackage[table]{xcolor} 
\usepackage{tabularx}

\begin{document}

%\title{Mechanical feedback regulates cell motility in growing tissues: From sub-diffusion to hyper-diffusion }
\title{Proliferation-driven mechanical feedback regulates cell dynamics in growing tissues}
\author{Sumit Sinha$^1$, Xin Li$^2$, Abdul N Malmi-Kakkada$^3$, and D. Thirumalai$^{1,2}$}
%\author{Sumit Sinha$^1$, Xin Li$^2$, D. Thirumalai$^2$}
%\email{dave.thirumalai@gmail.com}
\affiliation{$^1$Department of Physics, University of Texas at Austin, Austin, TX 78712, USA.}
\affiliation{$^2$Department of Chemistry, University of Texas at Austin, Austin, TX 78712, USA.} 
\affiliation{$^3$Department of Physics and Biophysics, Augusta University, Augusta, GA 30912, USA.}

\date{\today}

\begin{abstract}
 Local stresses in a tissue, a collective property, regulate cell division and apoptosis. In turn, cell growth and division induce active stresses in the tissue. As a consequence, there is a feedback between cell growth and  local stresses. However, how the cell dynamics depend on local stress-dependent cell division  and the feedback strength is not fully understood.  Here, we probe the consequences of stress-mediated growth and cell division on cell dynamics using  agent-based simulations of a two-dimensional growing tissue. We discover a rich dynamical behavior of individual cells, ranging from jamming (mean square displacement, $\Delta (t) \sim t^{\alpha}$ with $\alpha$ less than unity), to hyperdiffusion ($\alpha > 2$) depending on cell division rate and the strength of the mechanical feedback. Strikingly, $\Delta (t)$  is determined by the tissue growth law, which quantifies cell proliferation (number of cells $N(t)$ as a function of time). The growth law ($N(t) \sim t^{\lambda}$ at long times)  is regulated by the critical pressure that controls the strength of the mechanical feedback and the ratio between cell division-apoptosis rates. We show that $\lambda \sim \alpha$, which implies that higher growth rate leads to a greater degree of cell migration. 
 The variations in cell motility are linked to the emergence of highly persistent forces extending over several cell cycle times.  Our predictions are testable using cell-tracking imaging techniques.

\end{abstract}

\begin{titlepage}
\maketitle
\end{titlepage}


\section{Introduction}
\label{introduction}
Cell growth, proliferation, and apoptosis are ubiquitous in biology, and play a crucial role in embryogenesis, tumorigenesis, and wound healing \cite{barres1992cell,lecuit2007orchestrating}. The breakdown of strict control between cell division and apoptosis rates could lead to fatal diseases like cancer \cite{weinberg2013biology}. In cancer metastasis, the cells develop migratory phenotype and invade the surrounding tissues and organs \cite{kumar2009mechanics}. Therefore, to understand the role of cell division and apoptosis numerous experiments have been performed both in two and three dimensions, which provide the time traces of cells \cite{puliafito2012collective,valencia2015collective,Han20NatPhys,Kim20BBRC}. The cell trajectories could be used to calculate dynamical properties of cells \cite{sinha2020self} that may be quantitatively compared with experiments \cite{Valencia15}. 

An interplay between cell division, apoptosis, and biomechanical feedback determines cell proliferation and the associated dynamics in an evolving tissue. For instance, a growing tissue exhibits a morphological transition, characterized by contrasting collective cell dynamics in the pre-and post-transition phases \cite{puliafito2012collective}. Cells in the pre-transition phase exhibit fluid-like behavior whereas those in the post-transition phase are more solid-like \cite{puliafito2012collective}. The morphological transition, resulting in the contrasting dynamics, was attributed to the microenvironment-dependent growth and proliferation of cells \cite{shraiman2005mechanical,puliafito2012collective}. The growth of cells in tissue depends on the local stresses, which in turn depend on the local growth rate. In other words, there is a feedback between local stress and cell growth, as was pointed out in a prescient study nearly two decades ago \cite{shraiman2005mechanical}. In addition to fluid and solid-like behavior, the dynamics could also show glassy behavior in confluent \cite{czajkowski2019glassy} and non-confluent tissues \cite{malmi2018cell}. How the mechanical feedback and cell division affects the observed dramatic variations in collective cell dynamics as the tissue grows is largely unknown. 

Previous studies that considered cell growth and division on the cell collective dynamics assumed that the birth rate of cells depends on its coordination number \cite{matoz2017cell}. However, recent experiments report that mere contact between cells may not be sufficient for inhibiting  mitosis in cells \cite{puliafito2012collective}.   Here, using an agent-based model introduced previously \cite{drasdo2005single, schaller2005multicellular, malmi2018cell, malmi2022adhesion, sinha2020spatially} in which the growth of a cell depends on the local pressure, we establish that the dynamics of cells is linked to the tissue growth law. We show that tissue growth is controlled by two parameters-(a) the critical pressure ($p_c$) and (b) the birth rate of cells ($k_b$), which are intrinsic properties of individual cells. The $p_c$ value determines the mechanical feedback strength \cite{Gniewek19PRL}. 

The central results of this work are: (a) Depending on the values of $p_c$ and $k_b$, cells exhibit subdiffusive (the mean-squared displacement,  $\Delta(t)\propto t^{\alpha}, \alpha \leq 1$), superdiffusive ($1< \alpha \leq 2$) or even hyperdiffusive ($\alpha > 2$) dynamics. On increasing the value of $p_c$, the cells transition from sub-diffusive to hyperdiffusive dynamics. Surprisingly, on decreasing $k_b$, the cells switch from sub to super-diffusive or super to hyper-diffusive dynamics. (b) The tissue growth law exhibits a power increase in time ($t)$,  $N(t)\propto t^{\lambda}$, where $N$ is the number of cells. Strikingly, the global growth law is a predictor of the single-cell dynamics. As $\lambda$ increases, so does $\alpha$ with $\alpha \sim \lambda$. (c) The emergence of persistent forces due to cell division that extends over several cell cycle times is the principal reason for the anomalous (super or hyper-diffusive) cell dynamics.  Our work provides a unifying framework for understanding origins of differing dynamical regimes (sub-diffusive \cite{czajkowski2019glassy}, diffusive \cite{matoz2017cell} and super-diffusive \cite{malmi2018cell}) in the collective movement of cells driven by mechanical feedback arising from apoptosis and division.% sub-diffusive \cite{czajkowski2019glassy} to diffusive \cite{matoz2017cell} to super-diffusive \cite{malmi2018cell}.

\section{Methods}
\label{methods}
We briefly explain the off-lattice agent-based computational model used to simulate the spatio-temporal dynamics of a two-dimensional (2D) growing tissue. The computational model is adopted from previous studies \cite{drasdo2005single, schaller2005multicellular, malmi2018cell, malmi2022adhesion, sinha2020spatially,sinha2020self,sinha2022mechanical,nerger20243d}. The cells are represented as interacting deformable disks with radius depending on local rules, which %This simplified assumption of representing cells as deformable disks has been made previously \cite{matoz2017cell}. 
 assume that cells grow stochastically, and divide upon reaching a critical mitotic size ($R_m$). 
The interaction between cells is the sum of elastic and adhesive forces. We also assume that the cells are moving in an overdamped environment in which the inertia is negligible and viscous forces are large compared to environmental fluctuations.

%\textbf{Details of the model:}
\textit{Forces:} %The radius of each cell changes with time. 
%Several physical properties, such as the radius, elastic modulus, membrane receptor and ligand 
%concentration, and adhesive interaction between, characterize each cell.   
The elastic (repulsive) force between two disks of radii $R_{i}$ and $R_{j}$ is modeled as,
\begin{equation}
\label{rep}
F_{ij}^{el}(t) = \frac{h_{ij}^{3/2}(t)}{\frac{3}{4}(\frac{1-\nu_{i}^2}{E_i} + \frac{1-\nu_{j}^2}{E_j})\sqrt{\frac{1}{R_{i}(t)}+ \frac{1}{R_{j}(t)}}},
\end{equation}
where $E_{i}$ and $\nu_{i}$, respectively, are the elastic modulus and
Poisson ratio of cell $i$. The overlap between the disks, if they interpenetrate without deformation, is $h_{ij}$, which 
is defined as $\mathrm{max}[0, R_i + R_j - |\vec{r}_i - \vec{r}_j|]$ with $|\vec{r}_i - \vec{r}_j|$ being the center-to-center distance between the two disks. 
%\floatsetup[figure]{style=plain,subcapbesideposition=top}
%\begin{figure}
%\sidesubfloat[]{\includegraphics[width=0.7\linewidth] {cellcellinter.eps}\label{cellcellinter}}
%	\par\bigskip
%\sidesubfloat[]{\includegraphics[width=0.70\linewidth] {forcejkr.eps}\label{forcejkr}} 
%\caption{\textbf{(a)} Illustration of two interpenetrating cells $i$ and $j$ with radii $R_{i}$ and $R_{j}$, respectively.  The distance between the centers of the two cells is $|{\mathbf {r}}_{i} -{\mathbf {r}}_{j}|$, and their overlap is $h_{ij}$.
%\textbf{(b)} Force on cell $i$ due to $j$, ${\mathbf F}_{ij}$, for $R_{i}=R_{j}=4~\mu m$ using mean values of elastic modulus, poisson ratio, receptor and ligand concentration (see Table I). ${\mathbf F}_{ij}$ is plotted 
%as a function of distance between the centers of the two cells. 
%Inset shows the region where ${\mathbf F}_{ij}$ is attractive. When $|{\mathbf {r}}_{i} -{\mathbf {r}}_{j}|~\ge~R_{i}+R_{j}=8~\mu m$ the cells are no longer in contact, and hence, ${\mathbf F}_{ij}=0$.}
%end{figure}

\begin{figure}[h]
\includegraphics[width=15cm]{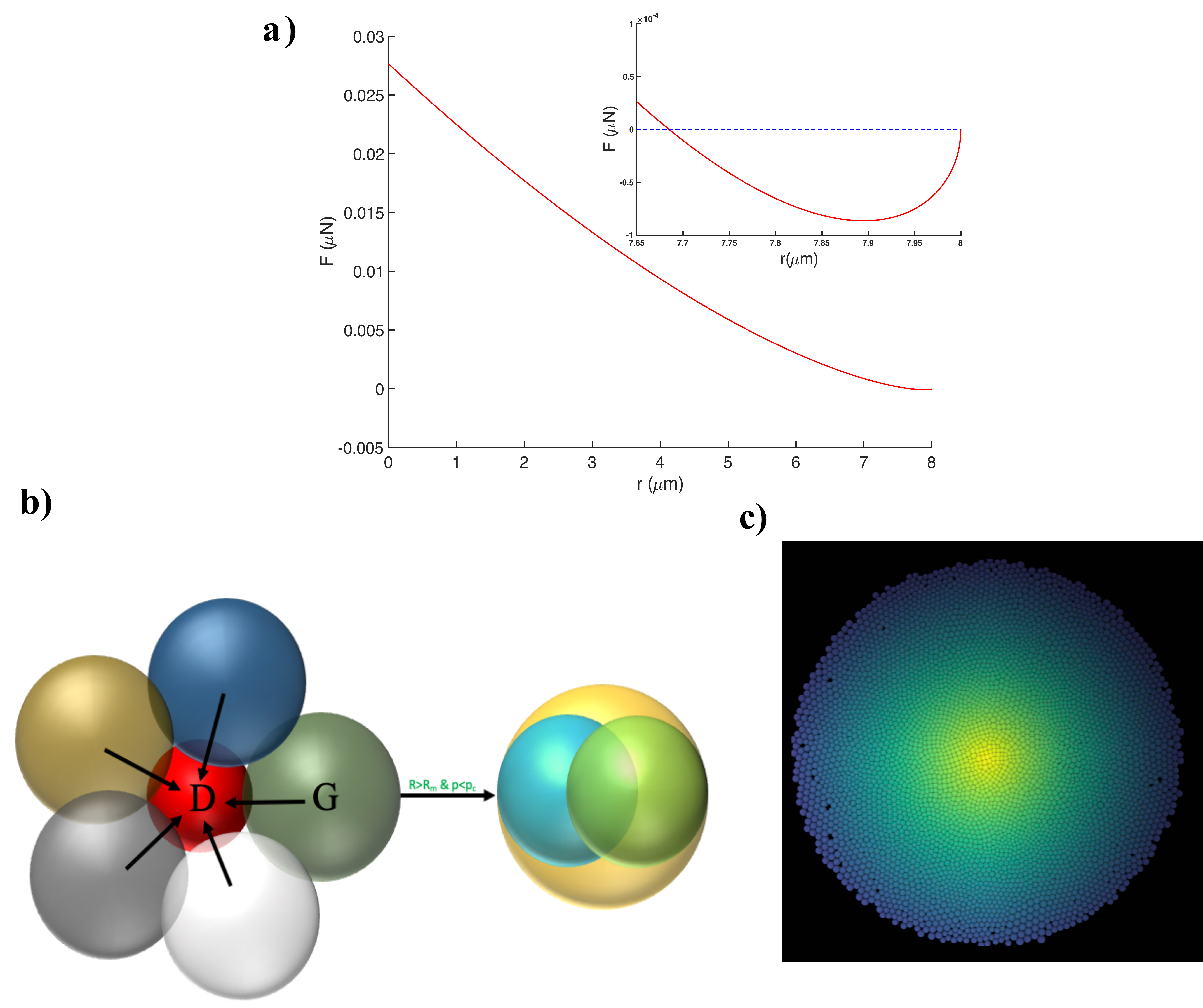}
\caption{\textbf{The 2D model.} (a) Inter-cellular force as a function of distance between two cells with identical radii, $R_i = R_j = 4 \mu m$. The repulsive and attractive parts of the force are given by Eqs. (1) and (2), respectively. The inset is the zoomed-in view that highlights the region in which the force is predominantly attractive. (b) Illustration of the role of mechanical feedback. On the left, the ``red” cell is dormant (cannot grow and divide) because the pressure exerted by the neighbors exceeds $p_c$. The ``green” cell is in the growth phase (G) ($p < p_c$). The green cell from the left gives birth to two daughter cells (cyan and green) when the radius exceeds the mitotic radius $R_m$. (c) A snapshot of the 2D growing tissue consisting of approximately 4,750 cells at $t^* =3.74$, with $p_c=10^{-3}MPa$ and $\frac{k_b}{k_a}=20$. The global shape  is approximately circular. }
\label{schematic}
\end{figure}

Cell adhesion, mediated by receptors on the cell surface, enables the cells to stick together. 
For simplicity, we assume that the receptor and ligand molecules are evenly 
distributed on the cell surface. Consequently, the magnitude of the attractive adhesive force, $F_{ij}^{ad}$, 
between two cells $i$ and $j$ scale as a function of their contact line segment, 
$L_{ij}$. Keeping the 3D model as a guide \cite{malmi2018cell}, we calculate $F_{ij}^{ad}$ using,
\begin{equation}
\label{ad}
F_{ij}^{ad} = L_{ij}f^{ad}\frac{1}{2}(c_{i}^{rec}c_{j}^{lig} + c_{j}^{rec}c_{i}^{lig}),
\end{equation}
where the $c_{i}^{rec}$ ($c_{i}^{lig}$) is the receptor (ligand) concentration 
(assumed to be normalized to the maximum receptor or ligand concentration so that  
$0 \leq c_{i}^{rec},  c_{i}^{lig} \leq 1$). The coupling constant $f^{ad}$ allows us to 
%\leq c_{i}^{(rec/lig)/max} 
rescale the adhesion force to account for the variabilities in the maximum densities of the receptor and ligand concentrations.  
We calculate the contact length, $L_{ij}$, using the length of contact between two intersecting circles,  $L_{ij} = \frac{\sqrt{(|4r_{ij}^2R_i^2-(r_{ij}^2-R_j^2+R_i^2)^2|)}}{r_{ij}}$. Here, $r_{ij}$ is the distance between cells $i$ and $j$. As before, $R_i$ and $R_j$ denote the radius of cell $i$ and $j$. In the present case, the strength of repulsive interactions is very large compared to attractive forces which can be seen in Figure \ref{schematic}a.

The  the sum of the repulsive and adhesive forces in Eqs.(\ref{rep}) and (\ref{ad}) point 
along the unit vector ${\bf n}_{ij}$ from the center of cell $j$ to the center of cell $i$. 
%The force exerted by cell $j$ on cell $i$, ${\bf F}_{ij}$, is shown in figure ~\ref{force_distance}, with the positive part denoting the repulsive regime and negative part depicting the attractive regime.
The total force on the $i^{th}$ cell is given by the sum over its nearest neighbors ($NN(i)$), 
\begin{equation}
{\bf F}_{i} = \Sigma_{j \epsilon NN(i)}(F_{ij}^{el}-F_{ij}^{ad}){\bf n}_{ij}. 
\end{equation}
The nearest neighbors satisfy the condition $R_i + R_j - |{\bf r}_i - {\bf r}_j|~>~0$.
%The force experienced by each cell is calculated only 

%\section{Simulations}

%\textbf{Equations of Motion:} %Once the force is calculated, 
\textit{Equation of Motion:} We used overdamped dynamics of the motion of the $i^{th}$ cell. The 
 equation of motion is, 
\begin{equation}
\label{eqforce}
\dot{{\bf r}}_{i} = \frac{{\bf F}_{i}}{\gamma_i}.
\end{equation}
Here, $\gamma_i$ is the friction coefficient of the $i^{th}$ cell. We assume $\gamma_i$ to be equal to $cR_i(t)$, where $c$ is a constant. Note, we neglect temperature effects because the drag forces are high \cite{matoz2017cell} compared to environmental fluctuations.

% \begin{figure}
%\begin{turn}{-90}
%\includegraphics[width=0.90\linewidth] {celldorgro.eps} %oa
%\end{turn}
%\caption{Cell dormancy (left panel) and cell division (right panel). If the local pressure $p_i$ that the $i^{th}$  cell experiences 
%(due to contacts with the neighboring cells) exceeds the critical pressure $p_{c}$, it enters 
%the dormant state ($D$).  Otherwise, the cells grow (G) until they reach the mitotic radius, $R_{m}$. 
%At that stage, the mother cell divides into two identical daughter cells with the same radius $R_{d}$. 
%We assume that the total volume upon cell division is conserved. A cell that is dormant at a given time can transit from that state at subsequent times.}
%\label{celldorgro}
%\end{figure}
\textit{Cell growth, division, and apoptosis:}
In the model, cells are either dormant ($D$) or in the growth ($G$) phase depending on the magnitude of the local pressure of the cell (see Figure \ref{schematic}b for a schematic).
Using Irving-Kirkwood's definition, we calculate the pressure ($p_i$) on the $i^{th}$ cell due to contact with its neighbors \cite{yang2014aggregation} using,
\begin{equation}
\label{pressure}
p_{i} =  \frac{1}{2}\Sigma_{j \epsilon NN(i)} \frac{{\bf F}_{ij} \cdot {\bf dr}_{ij}} {A_i},
\end{equation}
where $A_i$ is local area of influence, equal to $\theta \pi R_i^2$. We used $\theta=1.5$, in our simulations.  If the local pressure on the $i^{th}$ cell, $p_{i}$, exceeds a critical value ($p_c$) the cell ceases to grow and enters the dormant phase. Note that the cell can switch to the growth phase if $\frac{p_i}{p_c}<1$ at a later time. The critical pressure, $p_c$,  serves as a mechanical feedback  \cite{shraiman2005mechanical}. The local pressure, $p_i$, can easily exceed $p_c$ if it is small. In this case, most cells would be dormant for a long time. In the opposite limit, $p_c \gg p_i $, it is unlikely that the cells would reach the dormant phase. This would result in cell proliferation. Thus, $p_c$ is the strength of the mechanical feedback. A previous study used $p_c$ to control cell growth in confined spaces in a different context \cite{Gniewek19PRL}. It was shown there is a growth-driven jamming transition, controlled by the strength ($\propto \frac{1}{p_c}$) of the mechanical feedback. They did not consider cell dynamics, which is the focus of our investigation.  %The cell is considered to be in the dormant state. 

For growing cells, we assume that the area increases at a constant rate $r_A$ as the cell cycle progresses. 
The cell radius is updated from a Gaussian distribution with the mean rate $\dot{R} = (2\pi R)^{-1} r_A$.  
Over the cell cycle time $\tau$, $r_A$ is taken to be, 
\begin{equation}
r_A = \frac{\pi (R_{m})^2}{2\tau},
\end{equation}
where $R_{m}$ is the mitotic radius. The cell cycle time  is related to the growth rate ($k_b$) by $\tau=\frac{ln~2}{k_b}$. 
A cell divides once it grows to the fixed mitotic radius ($R_m$). 
To ensure the total area of a cell is conserved  upon cell division, we use 
$R_d = R_{m}2^{-1/2}$ as the radius of the daughter cells. The mother and daughter  cells are placed at a center-to-center distance, 
$d = 2R_{m}(1-2^{-1/2})$ upon cell division. The direction of the new cell location 
is chosen randomly from a uniform distribution on the unit circle. 
One source of stochasticity in the cell movement  is the random choice for  
the mitotic direction. The cells can also undergo apoptosis at rate $k_a$. In all the simulations, we vary $k_b$ but the apoptosis rate ($k_a$) is fixed to $10^{-6} s^{-1}$.
The values of the parameters used in the simulations are given in Table 1.

%\textbf{Initial Conditions:} 
We initiated the simulations by placing 100 cells on a 2D plane whose coordinates are chosen from a normal distribution with zero mean, and standard deviation $25~\mu m$. All the parameters except $p_c$ and $k_b$ are fixed. All the simulations are terminated when the scaled time $t^{*}=(k_b-k_a)t\sim 3.74$. A representative snapshot of the growing tissue is shown in Figure~\ref{schematic}c.

\textbf{Table I:}
The parameters used in the simulations. \par \bigskip

\begin{tabular}{ |p{7cm}||p{4cm}|p{5cm}|p{3cm}|  }
 \hline
 \bf{Parameters} & \bf{Values} & \bf{References} \\
 \hline
 Timestep ($\Delta t$)& 10$\mathrm{s}$  & This paper \\
 \hline
Critical Radius for Division ($R_{m}$) &  5 $\mathrm{\mu m}$ & ~\cite{schaller2005multicellular, malmi2018cell}\\
 \hline
Friction coefficient ($\frac{\gamma_i}{R_i}$) & 0.0942 $\mathrm{kg/ (\mu m~s)}$   & This paper  \\
 \hline
 Cell Cycle Time ($\tau_{min}$)  & 54000 $\mathrm{s}$  & ~\cite{freyer1986regulation, casciari1992variations,landry1981shedding,malmi2018cell}\\
 \hline
 Adhesive Coefficient ($f^{ad})$&  $10^{-4} \mathrm{\mu N/\mu m}$  & This paper \\
 \hline
Mean Cell Elastic Modulus ($E_{i}) $ & $10^{-3} \mathrm{MPa}$  & ~\cite{galle2005modeling,malmi2018cell}    \\
 \hline
Mean Cell Poisson Ratio ($\nu_{i}$) & 0.5 & ~\cite{schaller2005multicellular,malmi2018cell}  \\
 \hline
 Death Rate (\textcolor{black}{$k_a$}) & $10^{-6} \mathrm{s^{-1}}$ & ~\cite{malmi2018cell} \\
 \hline
Mean Receptor Concentration ($c^{rec}$) & 1.0 (Normalized) & ~\cite{malmi2018cell} \\
\hline
Mean Ligand Concentration ($c^{lig}$) & 1.0 (Normalized) & ~\cite{malmi2018cell}  \\
\hline
%\label{table1}
%\caption{Parameters used to carry out the simulations.}
\end{tabular}

\section{Results}
\label{result}
\subsection{Increasing $p_c$ with $\frac{k_b}{k_a}$ fixed enhances cell motility}
\label{pc_motion}
Local stress regulates cell division propensity, and hence, should influence the cell dynamics in a growing tissue. To assess the effect of feedback on cell dynamics, we varied $p_c$, strength of the feedback.  The dynamics are probed using the mean squared displacement ($\Delta (t)$),
\begin{equation}
  \Delta(t)=\frac{1}{N}\sum_{i=0}^{i=N}[{\bf r}_i(t)-{\bf r}_i(0)]^2 , 
\end{equation}
where ${\bf r}_i(t)$ is the position of the $i^{th}$ cell at time $t$, and $N$ is the number of cells. We also calculated the  tissue boundary, $\Delta r (t) $, an estimate of the tissue size, using, 
\begin{equation}
    \Delta r(t)=\frac{1}{N_b(t)}\sum_{i=1}^{N_b}|{\bf r}_i(t)-{\bf R}(t)|
\end{equation}
where $N_b(t)$ is the total number of boundary cells at time $t$ and ${\bf R}(t)$ is the center of the tumor at time $t$. These quantities can be readily measured using imaging experiments \cite{puliafito2012collective,valencia2015collective}. 

Figure \ref{changing_pc} shows the time dependence of  $\Delta(t)$ for three values of $p_c: 10^{-5} Nm^{-1}, 10^{-4} Nm^{-1}$ and $10^{-3} Nm^{-1}$ for a fixed $\frac{k_b}{k_a}=20$.  Because the cells undergo apoptosis,  we included only the cells that were present throughout the simulations in calculating $\Delta(t)$. In the intermediate time limit,  $t<\frac{1}{k_b-k_a}$, the dynamics is subdiffusive ( $\Delta(t)\sim t^{\delta},\delta <1$) for  the three $p_c$ values. The long time  ( $t>\frac{1}{k_b-k_a}$) dynamics depends on the $p_c$. We find that $\Delta(t)\sim (t^*)^{\alpha}$ is subdiffusive ($\alpha=0.68$)  for $p_c=10^{-5} Nm^{-1}$, superdiffusive ($\alpha=1.36$) for $p_c=10^{-4} Nm^{-1}$ and hyperdiffusive ($\alpha=3$) for $p_c=10^{-3} Nm^{-1}$. As the mechanical feedback strength increases, which is realized by decreasing $p_c$, the cells are jammed, resulting in slow dynamics at small $p_c$. \textcolor{black}{Increased cell proliferation with weaker mechanical feedback (larger $p_c$)} gives rise to superdifussive or even hyperdiffusive dynamics. 

 The invasion distance increases algebraically with time, $\Delta r(t)\sim (t^{*})^{\xi}$,  (the time $t^* = (k_b-k_a)t$) where $\xi$ characterizes the tissue invasion propensity. Figure \ref{boun_changing_pc} shows  $\Delta r(t)$ for $p_c$ equal to $10^{-5} Nm^{-1}, 10^{-4} Nm^{-1}$ and $10^{-3} Nm^{-1}$ with $\frac{k_b}{k_a}=20$. We find (Figure \ref{boun_changing_pc}) that the growing tissue is maximally invasive for $p_c=10^{-3} Nm^{-1}$ similar to behavior of $\Delta(t)$. For $p_c=10^{-5} Nm^{-1}, \xi= 0.34$, for $p_c=10^{-4} Nm^{-1}, \xi= 0.68$ and  for $p_c=10^{-3} Nm^{-1}, \xi= 1.23$. 
We surmise from the behavior of $\Delta(t)$ and $\Delta r(t)$ that the tissue dynamics is enhanced upon increasing $p_c$ at a fixed value of $\frac{k_b}{k_a}$. This is because the probability that the cells are in the growth phase increases as $p_c$ increases.

\subsection{Decreasing $\frac{k_b}{k_a}$ with $p_c$ fixed promotes cell migration}
\label{ratio_motion}
We next varied the cell division rate ($k_b$) at a fixed $k_a=10^{-6} s^{-1}$. 
%We again characterized the dynamics of the growing monolayer using $\Delta(t)$ and $\Delta r(t)$. 
Figure \ref{changing_ratio} shows $\Delta(t)$ as a function of $t$ for $\frac{k_b}{k_a}= 20, 8$ and $2$ at a fixed $p_c=10^{-4} Nm^{-1}$ .  Surprisingly, in the long time ($[k_b-k_a]t>1$) limit, slower dividing cells have higher  motility. %For $t>\frac{1}{k_b-k_a}$, $\Delta(t)\sim t^{\alpha}$. 
For instance,  $\frac{k_b}{k_a} = 20$, the MSD exponent values ($\Delta(t)\sim (t^*)^{\alpha}$ at long times),  are $\alpha=1.36$, for $\frac{k_b}{k_a}$ equal to 8, $\alpha=1.67$ and for $\frac{k_b}{k_a}$ equal to 2, $\alpha=2.90$.  We observed similar behavior for $\Delta r(t)$ on decreasing $\frac{k_b}{k_a}$. As before, we expressed $\Delta r(t)\sim (t^*)^{\xi}$. For $\frac{k_b}{k_a}$ equal to 20, $\xi=0.68$, for $\frac{k_b}{k_a}$ equal to 8, $\xi=0.85$ and for $\frac{k_b}{k_a}$ equal to 2, $\xi=1.23$. 

The time dependent changes in $\Delta r(t)$ and $\Delta (t)$ shows that the degree of migration, quantified using $\xi$ and $\alpha$, is enhanced upon decreasing $\frac{k_b}{k_a}$ as long as $p_c$ is fixed.

\subsection{Growth law dictates the dynamics of cells}
\label{g_law}
What is the unifying explanation for the non-trivial cell dynamics in an evolving cell colony as $p_c$ and $k_b$ are varied? The answer lies in the growth law of the tissue. The experimental growth law is determined by counting the number of cells as a function of time \cite{puliafito2012collective}. The growth law is an emergent property that depends not only on the properties  of individual cells but also the coupling, through the mechanical feedback and adhesive interactions.% to each other with the  being the dominant factor.

{\it Changing $p_c$}: We first calculated the number of cells ($N$) as a function of time   at  $p_c=10^{-5} Nm^{-1},10^{-4} Nm^{-1}$ and $10^{-3} Nm^{-1}$ with $\frac{k_b}{k_a}=20$ (Figure \ref{no_cells_pc}a). We find that $N(t)$ increases as, $N(t)\sim t^{\lambda}$. For $p_c=10^{-5} Nm^{-1}, \lambda=1$, for $p_c=10^{-4} Nm^{-1}$, $\lambda=1.31$ and for $p_c=10^{-3} Nm^{-1}$, $\lambda=2.78$. It is clear that growth rate  increases as the mechanical feedback strength  decreases (Figure \ref{no_cells_pc}a).  To determine the origin of the enhanced growth at as $p_c$ increases, we calculated the average pressure, $\langle p (t) \rangle= \frac{1}{N}\sum_{i=1}^{N}p_i$ (Figure \ref{no_cells_pc}b).   For $p_c=10^{-5}$, the average value of $\langle p (t) \rangle$ is higher than the critical pressure, which implies that the cells are predominantly in the dormant phase. For $p_c=10^{-4}$, $\langle p (t) \rangle$ exceeds $p_c$ after a few cell cycle times, and thus the cells start entering dormancy. However, for $p_c=10^{-3}$, $\langle p (t) \rangle$ is always smaller than $p_c$, which implies the majority of the cells are in the growth phase, resulting in \textcolor{black}{increased cell division, and proliferation. }

{\it Changing $\frac{k_b}{k_a}$}: We then calculated $N(t)$ at the fixed value of $p_c=10^{-4}$ for three values of $\frac{k_b}{k_a}=20,8$ and $2$.  Figure \ref{no_cells_pc}c shows $N(t)$ as a function of $\frac{k_b}{k_a}$.  The growth exponents ($N(t) \sim t^{\lambda}$) are $\lambda=1.31$, $\lambda=1.69$ and $\lambda=2.60$ for $\frac{k_b}{k_a}=20$, $\frac{k_b}{k_a}=8$, $\frac{k_b}{k_a}=2$, respectively. Strikingly, tissue growth rate decreases as cell division rate increases, which may be understood in terms of the dynamic changes in the average pressure, $\langle p (t) \rangle$, plotted in Figure \ref{no_cells_pc}d, as a function of  $\frac{k_b}{k_a}$. %The growth law explains the enhanced dynamics upon decreasing $\frac{k_b}{k_a}$. We also calculated $\langle p (t) \rangle$ (\ref{pres_changing_ratio}), which explains  the increase in growth as a function of  $\frac{k_b}{k_a}$. 
Upon decreasing $\frac{k_b}{k_b}$, the generation of pressure in the tissue is suppressed. For $\frac{k_b}{k_a}=2$, the $\langle p (t) \rangle$ is smaller than  $p_c$  for long times (exceeding the cell division time) unlike the case for $\frac{k_b}{k_a}=20$ and $8$. Therefore, multiples cell divisions occur in cells that divide slowly compared to those that divide fast, thus resulting in greater tissue growth. Our analyses show that for both conditions (changing $p_c$ and $\frac{k_b}{k_a}$), the growth law of the tissue determines the cell dynamics. %\textcolor{red}{ss: plot of growth exponent versus MSD - they roughly track each other it seems.}%\textcolor{red}{Sumit i is this not obvious?}

%\textcolor{blue}{The simulation results in Figure \ref{no_cells_pc} suggest that the cell dynamics is determined  the tissue growth law. The generality of this result follows from the following arguments.  If the overall shape of the tissue is circular in 2D (see Figure~\ref{schematic}c), we expect the exponents $\alpha$ ($\Delta (t) \sim t^{\alpha}$) and $\lambda$ ($N(t) \sim t^{\lambda}$) should have similar values with $\alpha \approx \lambda$. From the algebraic growth it follows that \textcolor{red}{$N(t) \sim t^{\lambda}\sim r^2$}, which holds for a circular shape. From the relation \textcolor{red}{$r^2 \sim \Delta (t) \sim t^{\alpha}$}, expect that $\alpha \sim \lambda$. In addition, the exponents $\xi$ ($\Delta r (t) \sim t^{\xi}$) and $\lambda$  should be related as $\lambda \approx 2\xi$.  The results in Figures~\ref{phase}a show that the relation is approximately satisfied.}
%, thus establishing that the 2D tissue is roughly circular.

\subsection{Emergence of highly correlated force}
\begin{figure}[h]
\includegraphics[width=18cm]{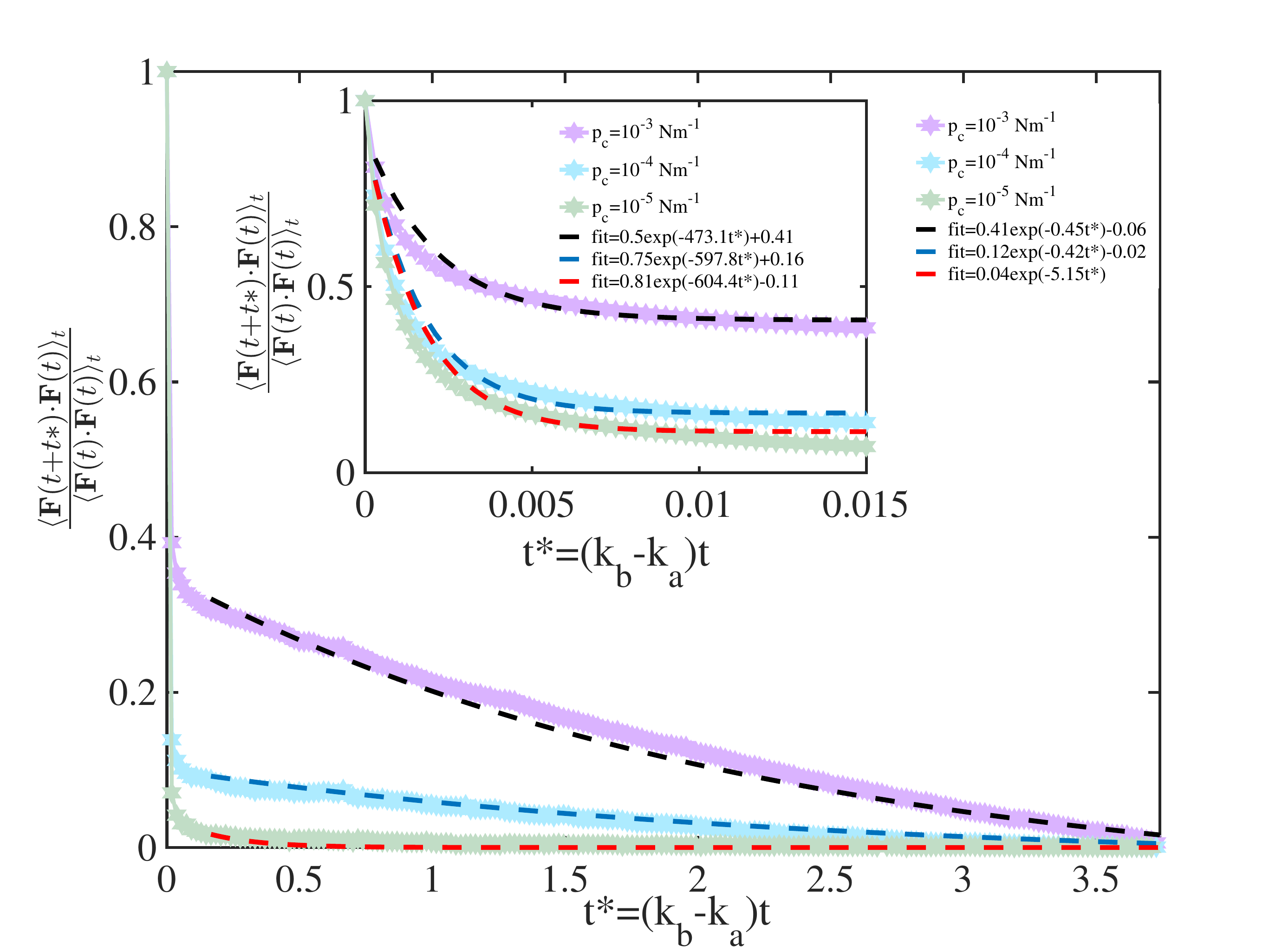}
\caption{\textbf{Correlation in force:} Force autocorrelation function (FAF) as a function of time. From top to bottom, FAF corresponds to $p_c=10^{-3}, 10^{-4}$ and $10^{-5}$. The dashed lines are the fits. Inset is the zoomed  of the initial times.  The figure shows the emergence of FAF with two-time scales:  long ($\sim \frac{1}{k_b-k_a}$) and short (elastic time scale $=\frac{\gamma}{ER}$).   }
\label{force_auto}
\end{figure}

To gain mechanistic understanding of the emergent anomalous dynamics of individual cells, we calculated the force autocorrelation function, FAF($t^{*}$) $=\frac{\langle{\bf F}(t+t^{*})\cdot {\bf F}(t)\rangle_t}{\langle{\bf F}(t)\cdot {\bf F}(t)\rangle_t}$ \cite{sinha2022mechanical}.  In an overdamped system, the FAF encodes the directed nature of motion in individual cells. Here, ${\bf F}(t)$ is the force on the cell at time $t$ and $\langle...\rangle_t$ is the time average. Figure \ref{force_auto} shows the plot of FAF for a fixed $\frac{k_b}{k_a}=20$ for $p_c=10^{-3} Nm^{-1}, 10^{-4} Nm^{-1}$ and $10^{-5} Nm^{-1}$. It shows that the FAF decays via a two steps, characterized by short ($\frac{\gamma}{ER_m}$) and  long ($\sim \frac{1}{k_b-k_a}$) times. To extract the two-time scales, we fit FAF using $Ae^{-\frac{t^*}{\tau_c}}+C$ in both the regimes. 

At short times (see the inset of Figure \ref{force_auto}), for $p_c=10^{-3} Nm^{-1}$, $A=0.5, \tau_c=\frac{1.2\gamma}{ER_m}$ and $C=0.42$. For $p_c=10^{-4} Nm^{-1}$, $A=0.75, \tau_c=\frac{0.97\gamma}{ER_m}$ and $C=-0.02$. Lastly, for $p_c=10^{-5} Nm^{-1}$, $A=0.81, \tau_c=\frac{0.95\gamma}{ER_m}$ and $C=0.1$. It is clear that at short times, the relaxation time is approximately close to the elastic time scale $\frac{\gamma}{ER_m}$, which is negligible compared  to $\frac{1}{k_b-k_a}$.

In the long time limit,  the FAF exhibits correlations.  For $p_c=10^{-3} Nm^{-1}$,  $A=0.4, \tau_c=\frac{2.2}{k_b-k_a}$ and $C=-0.06$. For $p_c=10^{-4}$, $A=0.12, \tau_c=\frac{2.3}{k_b-k_a}$ and $C=-0.02$. Lastly, for $p_c=10^{-5} Nm^{-1}$, $A=0.04, \tau_c=\frac{0.2}{k_b-k_a}$ and $C=0.003$. For $p_c=10^{-5} Nm^{-1}$, $A$ is negligible, implying the absence of correlations force, which explains the observed subdiffusive dynamics. The value of  $A$ for $p_c=10^{-3} Nm^{-1}$ is four times larger than for $p_c=10^{-4} Nm^{-1}$. In addition, the FAF decays over (2-3) cell division times when the feedback strength is high. Larger magnitude of FAF in the long time regime leads to higher degree of migration for $p_c=10^{-3} Nm^{-1}$.

\subsection{Diagram of states}
The simulation results in Figure \ref{no_cells_pc} suggest that the cell dynamics is determined by the tissue growth law. The generality of this result follows from the following arguments.  If the overall shape of the tissue is circular in 2D (see Figure~\ref{schematic}c), we expect the exponents $\alpha$ ($\Delta (t) \sim t^{\alpha}$) and $\lambda$ ($N(t) \sim t^{\lambda}$) should have similar values with $\alpha \approx \lambda$. From the algebraic growth of the tissue,  it follows that $N(t) \sim t^{\lambda}\sim r^2$, which holds for a circular shape. From the relation $r^2 \sim \Delta (t) \sim t^{\alpha}$, expect that $\alpha \sim \lambda$. In addition, the exponents $\xi$ ($\Delta r (t) \sim t^{\xi}$) and $\lambda$  should be related as $\lambda \approx 2\xi$.  The results in Figures~\ref{phase}a show that the relation is approximately satisfied.

Based on the findings in Figures~\ref{phase}a we are able to predict a diagram of states as a function of $p_c$ and $\frac{k_b}{k_b}$. Recent works probing the effect of cell division and apoptosis have reported subdiffusive \cite{czajkowski2019glassy}, diffusive \cite{matoz2017cell}, and superdiffusive motion \cite{malmi2018cell}. However, the regime in which these values emerge is unclear. Furthermore, time traces of cell positions maybe be recorded using particle tracking techniques. In anticipation of such experiments, we characterized single-cell dynamics by calculating the mean squared displacement over a broad range of $p_c$ and $\frac{k_b}{k_b}$.  We extracted the $\alpha$ exponent in the long time limit. The value of $\alpha$ could be used to determine the nature of dynamics in the time regime of interest. %Subdiffusion corresponds to $\alpha \leq 1$, superdiffusion for $1<\alpha \leq 2$ and hyperdiffusive dynamics for $\alpha > 2$. 
%Recent works probing the effect of cell division and apoptosis have reported subdiffusive \cite{czajkowski2019glassy}, diffusive \cite{matoz2017cell}, and superdiffusive motion \cite{malmi2018cell}. However, the regime in which these values emerge is unclear.  
%Therefore, we quantified the dynamics of cells, by extracting $\alpha$ in long time regime ($[k_b-k_a]t \geq 1$) as a function of  $\frac{k_b}{k_a}$ and $p_c$. We choose $p_c$ and $\frac{k_b}{k_a}$ is that they directly affect the cell division in the model (see the methods section).  
Figure \ref{phase}b shows the two-dimensional diagram of states. Interestingly,  we observe all three regimes of motion, subdiffusive, superdiffusive, and hyperdiffusive, by varying $\frac{k_b}{k_a}$ and $p_c$.  

Figure \ref{phase}b reveals three interesting characteristics of cell dynamics driven by cell division and apoptosis: (a) Upon increasing $p_c$, thare is a transition from subdiffusive to superdiffusive, and finally hyperdiffusive diffusive behavior. At a fixed $\frac{k_b}{k_a}=20$, for $p_c=5\times 10^{-6} Nm^{-1}$ dynamics is subdiffusive, for $p_c=10^{-4} Nm^{-1}$ cells exhibit superdiffusive motion. Upon further increasing  $p_c$ to $10^{-3} Nm^{-1}$, hyperdiffusive dynamics is observed. (b) Surprisingly, upon decreasing $\frac{k_b}{k_a}$, $\alpha$ increases. For smaller (higher) $p_c$ values, on decreasing $\frac{k_b}{k_a}$, the dynamics change from subdiffusive (superdiffusive) to superdiffusive (hyperdiffusive) behavior. For fixed $p_c=10^{-5} Nm^{-1}$, at $\frac{k_b}{k_a}=20$ ($\frac{k_b}{k_a}=2$), we observe subdiffusion (superdiffusion). For a higher value of $p_c=10^{-4} Nm^{-1}$, at $\frac{k_b}{k_a}=20$ ($\frac{k_b}{k_a}=2$), the dynamics is superdiffusive ion (hyperdiffusive). The diagram of states (Figure \ref{phase}b)  was created using a smoothing procedure where the values of the MSD exponents at unknown values of $\frac{k_b}{k_a}$ and $p_c$  were interpolated using the known MSD values (obtained via simulations). The interpolation is logarithmicaly (linearly) scaled in $p_c$ ($\frac{k_b}{k_a}$) axis. 
The two-dimensional phase diagram predicts the emergence of different dynamical regimes, from subdiffusive to hyperdiffusive, which can be tested in imaging experiments \cite{puliafito2012collective, valencia2015collective}.

\section{Conclusion}
\label{discussion}
%Dynamics of cells in a growing tissue is an active area of research \cite{valencia2015collective, puliafito2012collective}. However, different theoretical models capture different aspects of dynamics  \cite{czajkowski2019glassy,matoz2017cell,malmi2018cell}.  %subdiffusive, diffusive \cite{matoz2017cell} to superdiffusive \cite{malmi2018cell}. 
Using a two-dimensional off-lattice model, we have provided a comprehensive picture of the variations in the dynamics as the strength of the mechanical feedback and cell division rates are altered.  The dynamics change from subdiffusive to superdiffusive to hyperdiffusive, as the $\frac{k_b}{k_a}$ and $p_c$ are varied.   We also showed that in growing tissue, highly persistent forces emerge as the strength of the mechanical feedback increases,  whose decay exhibits two relaxation time scales: \textcolor{black}{one short (elastic time scale, $\frac{\gamma}{ER_m}$) and one long (division-apoptosis time scale, $\frac{1}{k_b-k_a}$).} The presence of persistent forces determines variations in the dynamics as cell division rates and the strength of the feedback are varied. Strikingly, the cell dynamics are controlled by the growth law of the tissue, which depends primarily on the strength of the mechanical feedback. Interestingly, the three exponents $\alpha$, $\lambda$ and $\xi$ from cell dynamics and tissue growth are related as $\alpha \approx \lambda \approx 2\xi$. Therefore, we can estimate the values of the other two exponents if one of them is obtained in experiments. The phase diagram summarizing our findings provides a unified picture of the disparate dynamics found in several theoretical studies. \cite{czajkowski2019glassy,matoz2017cell,malmi2018cell}

\textbf{Acknowledgements:} This work was supported by a grant from the National Science Foundation (Phy 23-10639) and the Welch Foundation (F-0019).
%The individual characteristics that play a crucial role are critical pressure $p_c$ and ratio of division-apoptosis rates $\frac{k_b}{k_a}$.

%We also showed that in growing tissue, highly persistent forces emerge,  whose decay exhibits two relaxation time scales: \textcolor{black}{one short (elastic time scale, $\frac{\gamma}{ER_m}$) and one long (division-apoptosis time scale, $\frac{1}{k_b-k_a}$).} The presence of persistent forces determines variations in the dynamics as cell division rates and the strength of the feedback are varied.   

\clearpage

\clearpage
\floatsetup[figure]{style=plain,subcapbesideposition=top}
\begin{figure}

	\sidesubfloat[]{\includegraphics[width=0.5\textwidth]{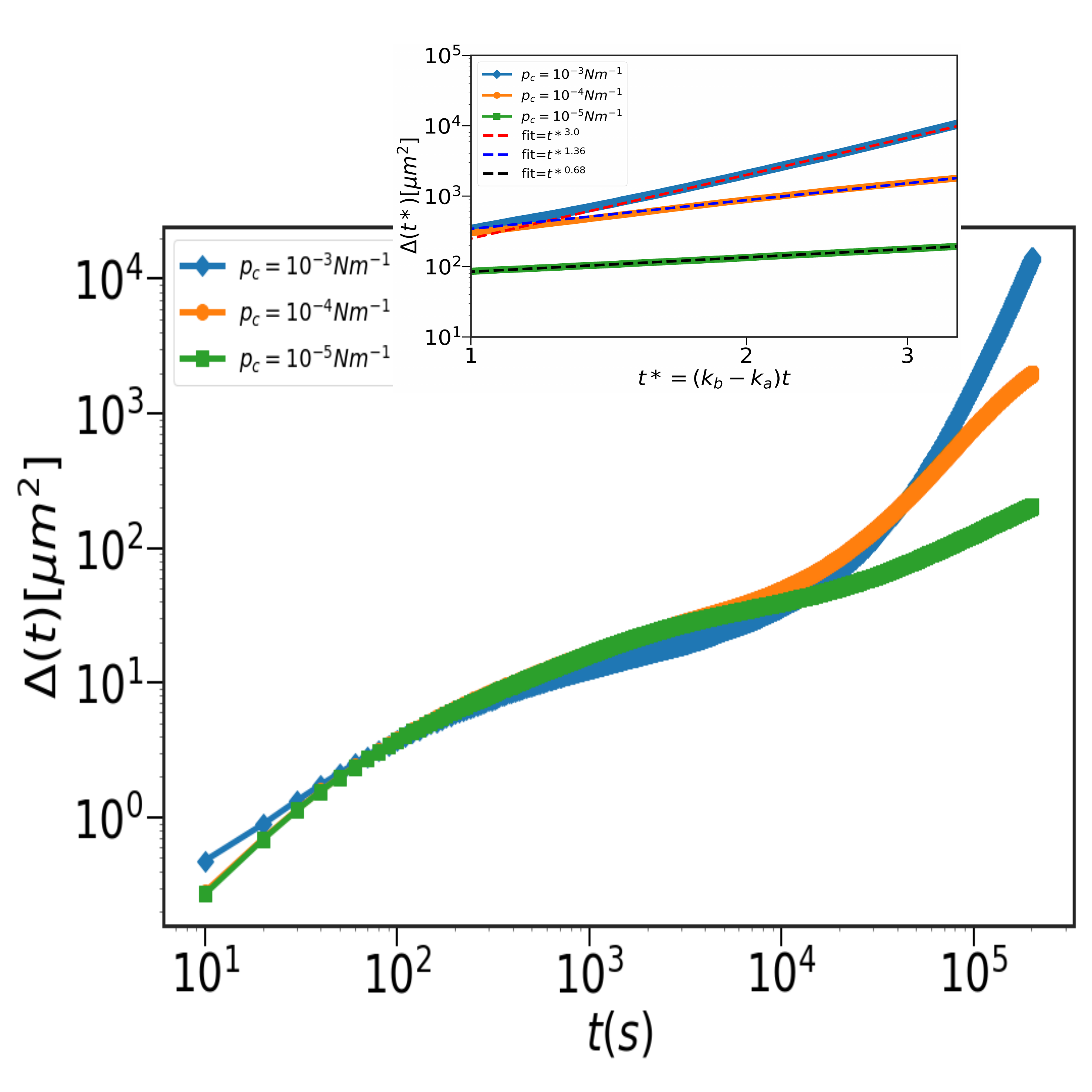} \label{changing_pc}}
	\sidesubfloat[]{\includegraphics[width=0.5\textwidth] {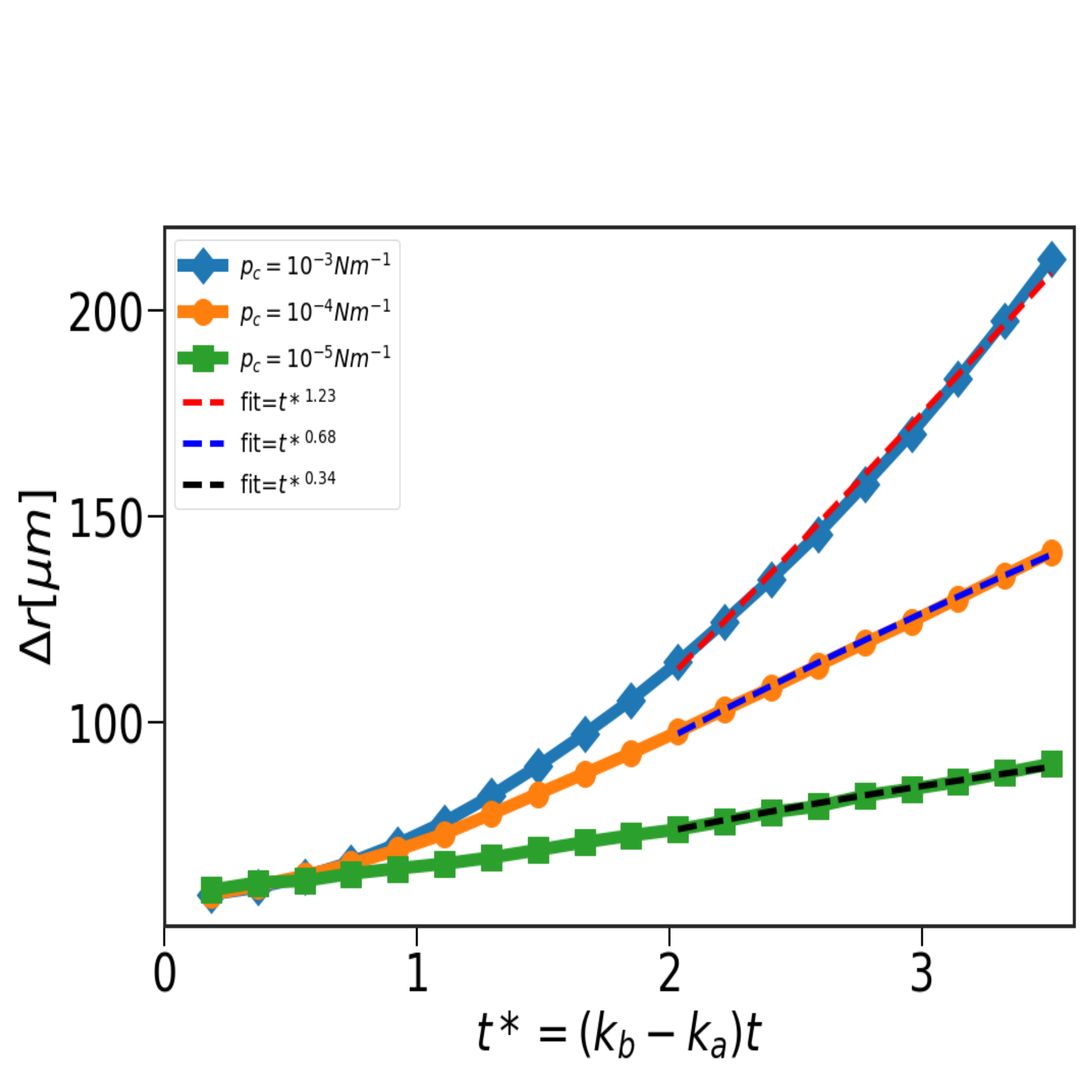} \label{boun_changing_pc} }
	\par
	\sidesubfloat[]{\includegraphics[width=0.5\textwidth] {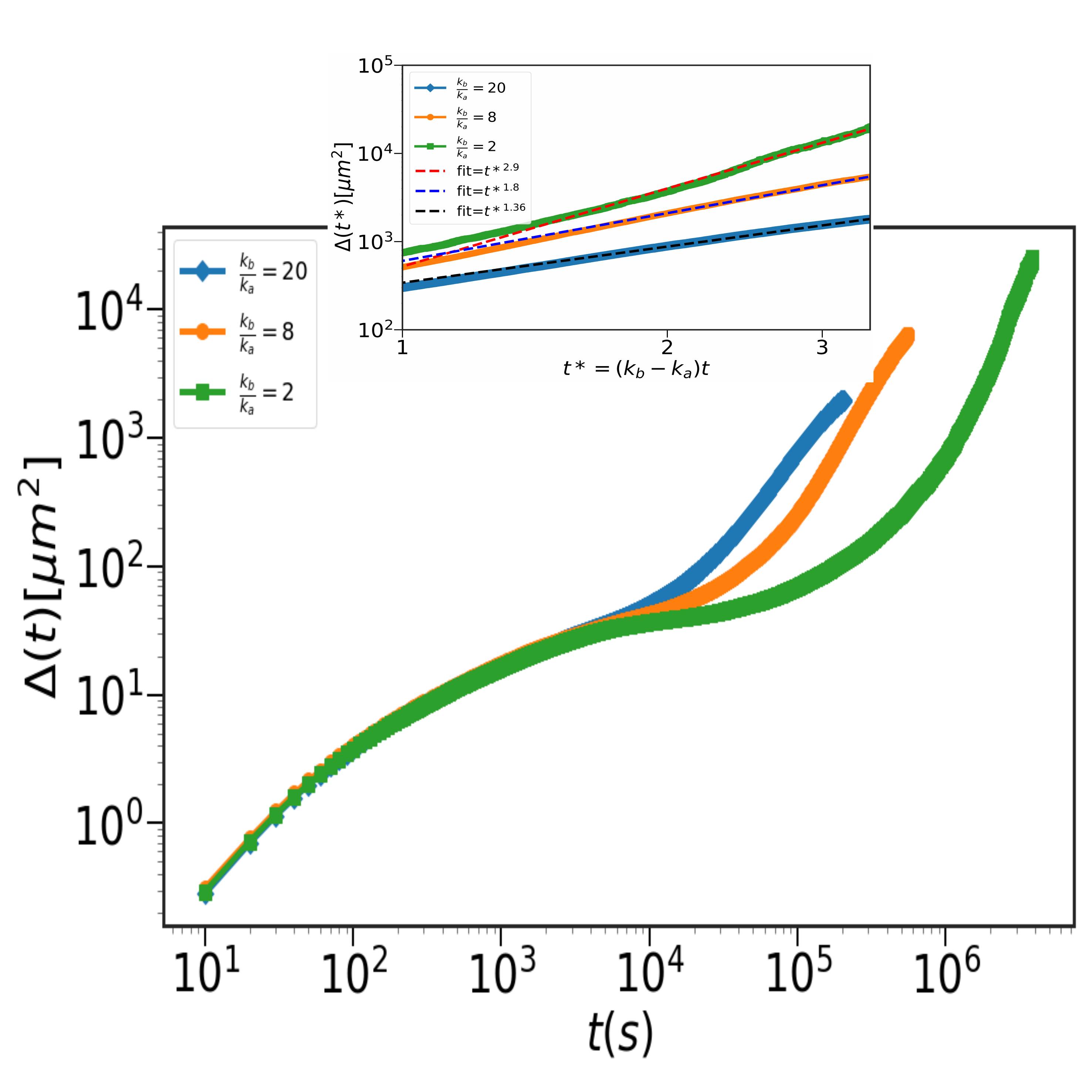} \label{changing_ratio} }
	\sidesubfloat[]{\includegraphics[width=0.5\textwidth] {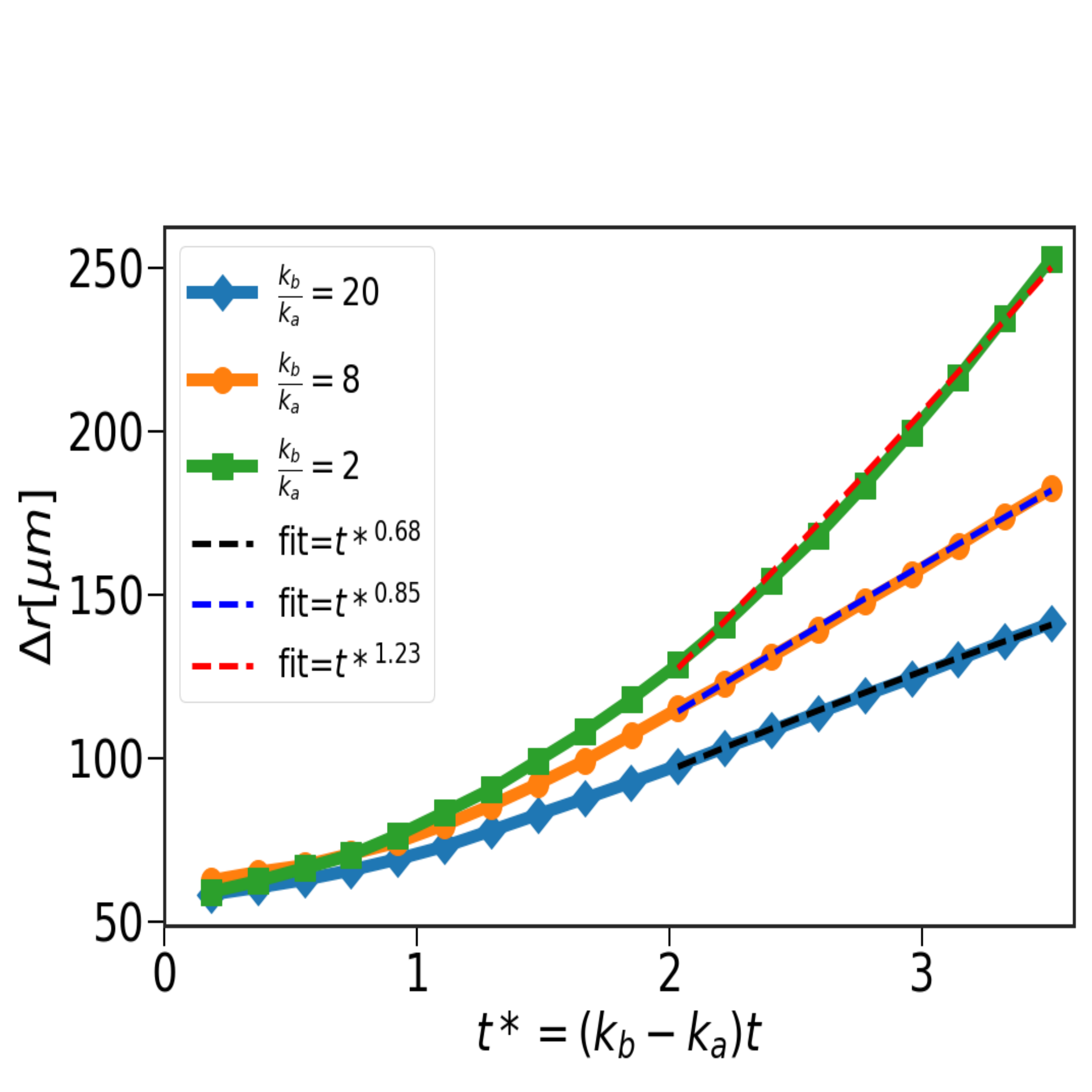} \label{boun_changing_ratio} }
\caption{\textbf{Cell dynamics is regulated by $p_c$ and $\frac{k_b}{k_a}$:} ({\bf a}) Mean squared displacement, $\Delta(t)$, as a function of time. From top to bottom, the curves are for $p_c=10^{-3} Nm^{-1}, 10^{-4} Nm^{-1}$ and $10^{-5} Nm^{-1}$. The inset focuses on the long time limit ($t>\frac{1}{k_b-k_a}$). The x-axis is scaled by $k_b-k_a$. The dashed lines are  power law fits ($\Delta(t)\sim t^{\alpha}$). The $\alpha$ values are given in the upper left box.  (Continued on the next page)} 
\label{msd_changing_pc_birth}
\end{figure}

\clearpage
\begin{figure}
\contcaption{({\bf b}) Invasion distance, $\Delta r (t)$ as a function of time for different $p_c$ values. The dashed lines are power-law fits ($\Delta r \sim (t^*)^{\xi}$). The $\xi$ values are given in the upper left box. ({\bf c}) $\Delta(t)$, as a function of time. From left to right, curves correspond to $\frac{k_b}{k_a}=20, 8$ and $2$. The inset focuses on the long time regime ($t>\frac{1}{k_b-k_a}$). The dashed lines are the power law fits ( $\Delta(t)\sim (t^*)^{\alpha}$). The $\alpha$ values are given in the upper left box. ({\bf d}) Invasion distance, $\Delta r (t)$ as a function of time for changing $\frac{k_b}{k_a}$. The dashed line is the power law fit ($\Delta r \sim (t^*)^{\xi}$). The $\xi$ values are given in the upper left box.}  
\end{figure}

\clearpage
\floatsetup[figure]{style=plain,subcapbesideposition=top}
\begin{figure}
\includegraphics[width=17cm]{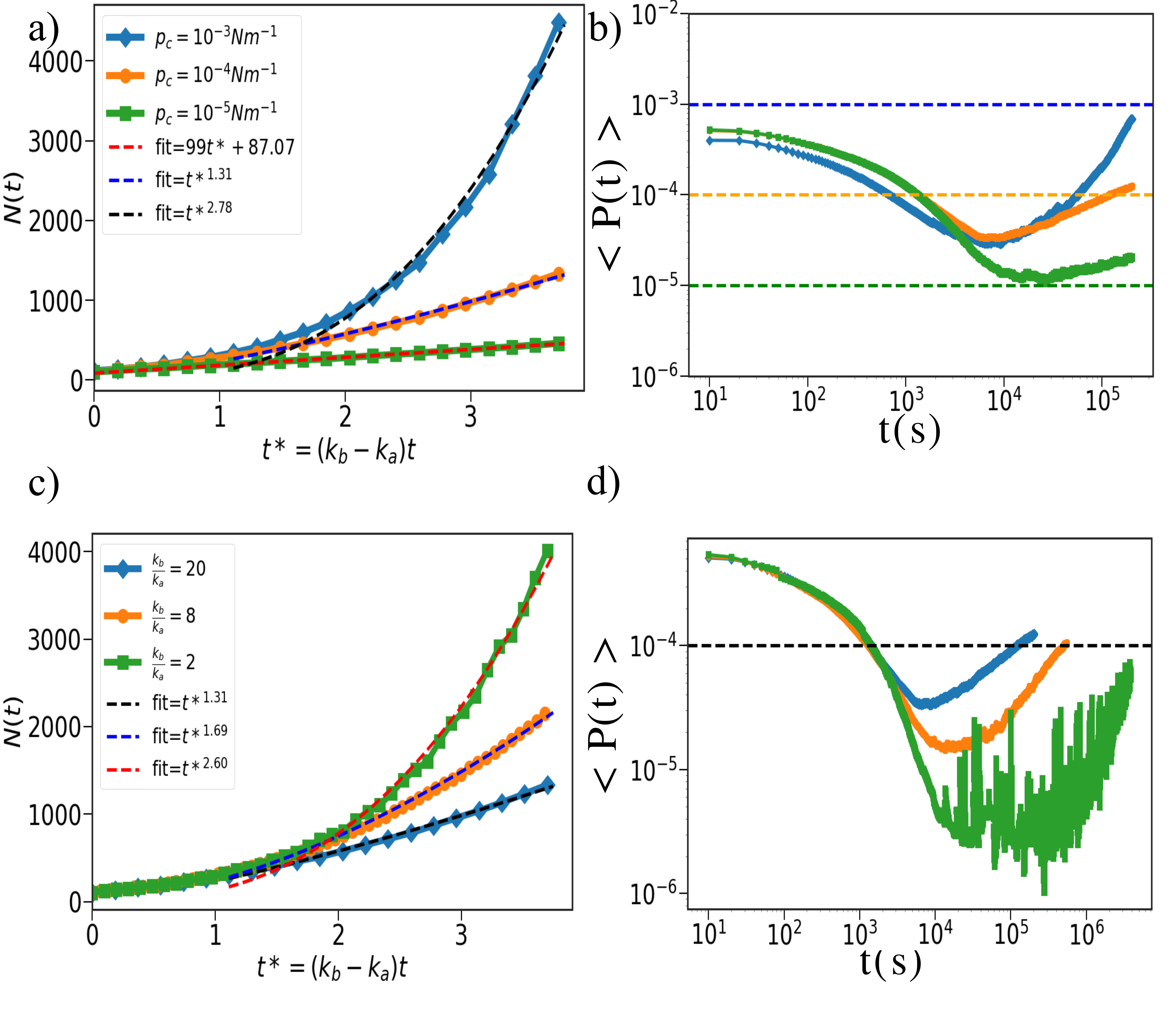}
\caption{ {\bf  Growth law governs the cell dynamics:} ({\bf a}) Number of cells, ($N(t)$), as a function of time at three values of $p_c$, labeled in the figure. 
The dashed lines with the power the power law fits ( $N(t)\sim (t^*)^{\lambda}$) are shown.  ({\bf b}) Average pressure, \textcolor{black}{$\langle P(t)\rangle$}, as a function of time. The curves correspond to $p_c=10^{-3} Nm^{-1} (top), 10^{-4} Nm^{-1}$ (middle), and $10^{-5} Nm^{-1}$ (bottom). The dashed lines mark the $p_c$ values;  blue - $p_c=10^{-3} Nm^{-1}$, orange - $ p_c=10^{-4} Nm^{-1}$,  and green - $ p_c=10^{-5} Nm^{-1}$. (Continued on the next page).}
\label{no_cells_pc}
\end{figure}

\clearpage
\begin{figure}
\contcaption{ ({\bf c}) $N(t)$, as a function of time. From bottom to top, curves correspond to $\frac{k_b}{k_a}=20$ (blue), $8$ (orange) and $2$ (green). The dashed lines are the power law fits. The $\lambda$ values are mentioned in the upper left box. ({\bf d}) Average pressure, $\langle P(t)\rangle$, as a function of time for the three $\frac{k_b}{k_a}$ values. From bottom to top, curves correspond to $\frac{k_b}{k_a}=20$ (blue), $8$ (orange) and $2$ (green). The dashed line corresponds to a pressure equal to $10^{-4} Nm^{-1}$.  }
\end{figure}

\clearpage

\clearpage
\begin{figure}[h]
\includegraphics[width=12.0cm]{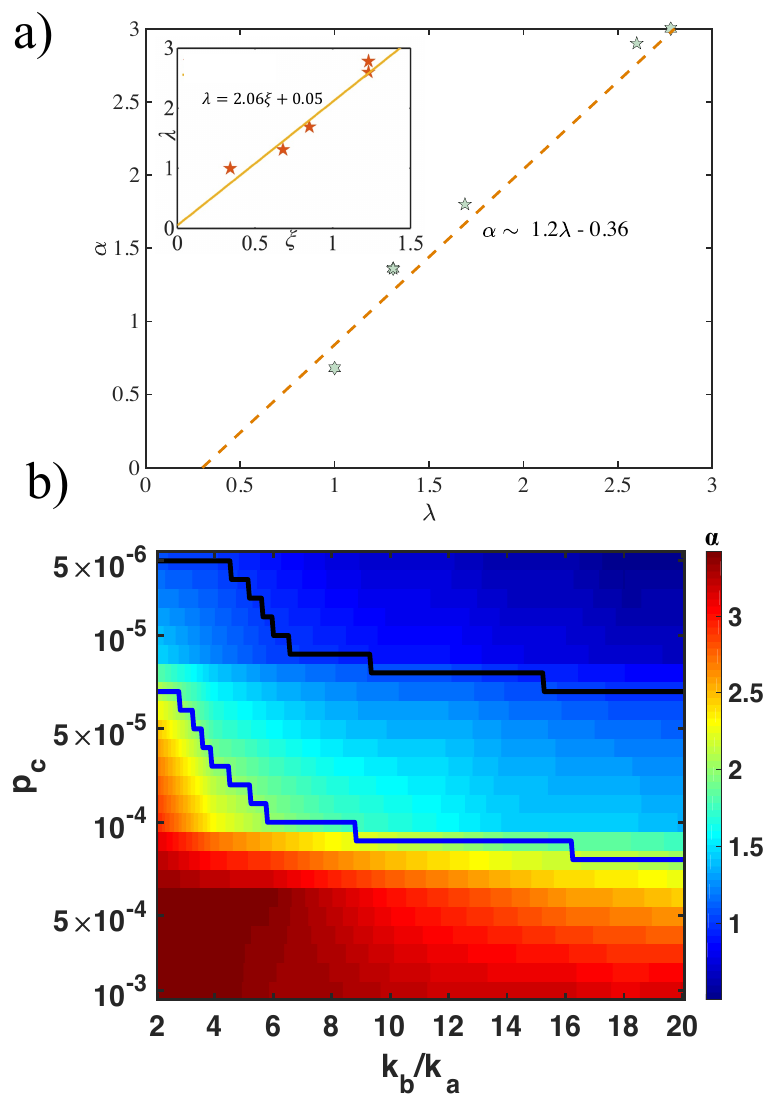}
\caption {\textbf{Dynamical phase diagram :} (a) The MSD exponent $\alpha$ as a function of the growth law exponent $\lambda$. The slope of the dashed line is approximately unity. In the inset we plot the relationship between $\lambda$ and $\xi$. The fit of the line is $\lambda \approx 2\xi$. (b) Dynamical regimes as a phase diagram in the plane of $p_c$ and  $\frac{k_b}{k_a}$.   The color bar on the right shows the value of $\alpha$.   Sub-diffusion ($\alpha \leq 1$), superdiffusion ($1<\alpha \leq 2$), and hyper-diffusion ($\alpha>2$) in the long-time cell dynamics ( $(k_b-k_a)t>1$). The black (blue) lines correspond to $\alpha=1$ ($\alpha=2$).  }
\label{phase}
\end{figure}

\clearpage
\begin{figure}    
    \contcaption{\textcolor{black}{The two-dimensional phase diagram predicts the emergence of subdiffusion, superdiffusion, and hyperdiffuison, depending on the values of $p_c$ and $\frac{k_b}{k_a}$. The phase diagram was obtained via a smoothing procedure (details in the text)}}
\end{figure}

\clearpage
\bibliographystyle{unsrt}
\bibliography{2D_paper.bib}

\end{document}